%% file: exclpi.tex
\documentclass[prl,nofootinbib,floatfix,twocolumn,showpacs,superscriptaddress]{revtex4}
\usepackage{graphicx,amsmath}

\def\hermes{{\sc Hermes~}}

\def\desy{{\sc Desy~}}
\def\hera{{\sc Hera~}}

\def\pythia6{{\sc Pythia6~}}
\def\pythia{{\sc Pythia~}}

\newcommand{\ut}{\text{UT},\ell}

\newcommand{\uu}{\text{UU}}

\begin{document}

\title{Single-spin azimuthal asymmetry in exclusive electroproduction of $\pi^{+}$ mesons on transversely polarized protons}

\input{authors}

\begin{abstract}
Exclusive electroproduction of $\pi^+$ mesons was studied 
by scattering $27.6\,\mathrm{GeV}$ positrons or   
electrons off 
a transversely polarized hydrogen target. 
The single-spin azimuthal asymmetry 
with respect to target polarization 
was measured as a function of the 
Mandelstam variable $t$, the Bjorken scaling variable $x_\text{B}$, 
and the virtuality $Q^2$  
of the exchanged photon. 
The extracted Fourier components of the asymmetry were found to be 
consistent with zero, except one that was found 
to be large and that involves interference of contributions 
from longitudinal and transverse virtual photons.

\end{abstract}


\maketitle

Generalized Parton Distributions (GPDs)~\cite{Muller,Rady,Ji} 
provide a three-dimensional representation of the nucleon structure 
at the partonic level correlating the longitudinal momentum
fraction of a parton with its transverse spatial
coordinates~\cite{Burkardt,Diehl2002,Ralston,Belitsky2002,Burkardt2003}.
The possibility to study GPDs relies on 
factorization theorems  
proven in the framework of perturbative 
quantum chromodynamics 
for hard exclusive processes
at leading twist, 
in particular for 
hard production of mesons  
by longitudinal 
virtual photons~\cite{Collins}. 
For recent theoretical reviews, see~\cite{Goeke,Diehl,Belitsky}.

In the description of 
hard exclusive electroproduction of pseudoscalar 
mesons at leading twist,    
only the two GPDs $\widetilde H$ and $\widetilde E$ appear.
Spin-averaged and spin-dependent cross sections are sensitive 
to different combinations of $\widetilde H$ and $\widetilde E$. 
It was predicted  
that for exclusive production of 
$\pi^+$ mesons on transversely polarized protons by longitudinal 
virtual photons the interference between the 
pseudovector ($\propto\widetilde H$) 
and 
pseudoscalar ($\propto\widetilde E$) contributions 
to the cross section 
leads to a large proton-spin related azimuthal 
asymmetry~\cite{Pobylitsa,Frankfurt}. 
Unlike the spin-averaged cross section, 
this asymmetry is directly proportional to 
the sine of the relative phase between $\widetilde H$ and $\widetilde E$.
It was shown 
that next-to-leading order corrections in the strong-coupling 
constant $\alpha_s$ cancel in the asymmetry~\cite{Belitsky2001,DiehlKugler}. 
No GPD-based model  
predictions are available for the production of 
$\pi^+$ mesons by transverse virtual photons as
no factorization theorems exist  
for this case, but 
also because 
the leading-twist contribution is expected to be dominant. 
Measurements of the asymmetry are considered 
to be a valuable source of information 
about possible contributions from transverse virtual 
photons~\cite{GK2009}. 
In a Fourier expansion of the proton-spin-dependent part of the 
hard exclusive pion electroproduction cross section~\cite{DiehlSapeta} 
the only leading-twist contribution to the asymmetry 
from longitudinal virtual photons  
is the $\sin(\phi-\phi_S)$ Fourier amplitude, which can be used to test 
GPD models. All other amplitudes involve contributions 
from transverse virtual photons.  
Here, following the Trento conventions~\cite{Trento}, 
$\phi$ and $\phi_S$ are the azimuthal angles in the proton rest frame
of the 
pion-momentum and the proton-polarization vectors, 
respectively, measured about the 
virtual-photon momentum vector 
relative to the lepton scattering plane. 
For recent theoretical analyses of exclusive pion 
electroproduction, see~\cite{GK2009,BM2009,KM2009}. 

The \hermes collaboration has previously  
performed measurements 
of the spin-averaged cross section~\cite{hermes:xsection} and the  
single-spin azimuthal asymmetry in exclusive $\pi^+$ 
electroproduction on longitudinally polarized protons~\cite{Hermes2002}. 
This letter 
reports the first measurement of the single-spin azimuthal asymmetry 
for the hard exclusive reaction $e p^{\uparrow}\to e n \pi^+$ on 
transversely polarized protons.  
The kinematic variables relevant for the analysis of this 
process are the squared four-momentum of the exchanged 
virtual photon $q^2\equiv -Q^2$, the Bjorken variable 
$x_\text{B}\equiv Q^2/(2M_p\nu)$, 
and the squared four-momentum transfer $t\equiv (q-p_{\pi^+})^2$. 
Here, $M_p$ is the 
proton mass, $\nu$ the energy of the virtual photon in the target rest 
frame, and $p_{\pi^+}$ the four-momentum of the pion. 
Instead of $t$, the quantity 
$t^\prime\equiv t-t_0$ is used in the analysis, where $-t_0$ represents the 
minimum value of $-t$ for a given value of $Q^2$ and $x_\text{B}$. 

The data corresponding to an integrated luminosity of 
$0.2\,\mathrm{fb}^{-1}$ 
were collected with the \hermes spectrometer~\cite{hermes:spectr} 
in the years $2002$-$2005$. The $27.6\,\mathrm{GeV}$ positron 
or electron beam was scattered off the transversely nuclear-polarized 
gaseous hydrogen target internal to the \hera storage ring at {\sc Desy}.  
The open-ended target cell was fed by an atomic-beam 
source~\cite{hermes:target1} 
based on Stern-Gerlach separation combined with 
radiofrequency transitions of 
hydrogen hyperfine states. The nuclear polarization of the atoms 
was flipped at $\mbox{1\text{-}3}\,\mathrm{minute}$ time intervals, while both 
this polarization and the atomic fraction inside the target cell were 
continuously measured~\cite{hermes:target2}.  
The average magnitude of the transverse polarization of the target 
with respect to the beam direction was 
$|P_\text{T}|=0.72\pm0.06$. 

Events were selected with exactly two tracks of 
charged particles: a lepton and a pion. Furthermore, it was required that 
no additional energy deposition was detected in the 
electromagnetic calorimeter. 
The \hermes geometrical acceptance of $\pm 170\,\mathrm{mrad}$  
horizontally and $\pm(40$-$140)\,\mathrm{mrad}$ vertically resulted  
in detected scattering angles ranging from $40\,\mathrm{mrad}$  
to $220\,\mathrm{mrad}$. Leptons 
were identified with an average efficiency of $98\%$ and a
hadron contamination of less than $1\%$ by using an electromagnetic 
calorimeter, 
a transition-radiation detector, a preshower scintillation counter, and 
a dual-radiator ring imaging \v{C}erenkov detector~\cite{hermes:rich}.  
Pions were identified in the momentum range 
$2\,\mathrm{GeV}<p<15\,\mathrm{GeV}$ using the \v{C}erenkov detector. 
For this momentum range the pion identification efficiency was on average 
$99\%$ and the contamination from other hadrons less than $2\%$. 
The kinematic requirement $Q^2>1\,\mathrm{GeV}^2$ was imposed on the scattered 
lepton in order to select the hard scattering regime.  

The single-spin asymmetry for exclusive $\pi^+$ production 
with unpolarized (U) beam and target polarization 
transverse (T) to the lepton ($\ell$) beam direction is defined 
as 
\begin{align}\label{eq1}
A_{\ut}(\phi,\phi_S)=
\frac{1}{|P_\text{T}|}
\frac{\text{d}\sigma^{\uparrow}(\phi,\phi_S)
-\text{d}\sigma^{\downarrow}(\phi,\phi_S)} 
{\text{d}\sigma^{\uparrow}(\phi,\phi_S)
+\text{d}\sigma^{\downarrow}(\phi,\phi_S)}, 
\end{align}
where 
$\text{d}\sigma^{\uparrow(\downarrow)}(\phi,\phi_S)=
\text{d}\sigma_{\uu}(\phi)+P_\text{T}\,\text{d}
\sigma_{\ut}(\phi,\phi_S)$ 
is a sum of 
the spin-averaged and spin-dependent cross sections, 
with $P_\text{T}/|P_\text{T}|$ equal to $1$ $(-1)$ for
the $\uparrow$ ($\downarrow$) orientations
of the transverse target polarization vector $\text{\boldmath$P$}_\text{T}$.
Both numerator and denominator 
of (\ref{eq1}) 
can be Fourier-decomposed~\cite{DiehlSapeta}, respectively, as 
\begin{align}\label{eq1a}
\text{d}\sigma_{\ut}(\phi,\phi_S)\propto 
2\langle\sin(\phi-\phi_S)\rangle_{\ut}\sin(\phi-\phi_S)+\dots, 
\end{align}
where the ellipsis denotes five more terms omitted here for brevity, 
and 
\begin{align}\label{eq1c}
\begin{split}
\text{d}\sigma_{\uu}(\phi)\propto 1 &+2\langle\cos\phi\rangle_{\uu}\cos\phi \\
                            &+2\langle\cos(2\phi)\rangle_{\uu}\cos(2\phi).
\end{split}
\end{align}
Ideally, the Fourier amplitudes in (\ref{eq1a}), 
which provide most direct access to the photoabsorption subprocesses, 
should be measured, e.g., 
\begin{align}\label{eq1b}
\begin{split}
&\langle \sin(\phi-\phi_S) \rangle_{\ut} \\
&=
\frac{\int\!{\text{d}\phi \text{d}\phi_S}\,\sin(\phi-\phi_S)\,\text{d}\sigma_{\ut}(\phi,\phi_S)}
{\int\!{\text{d}\phi \text{d}\phi_S}\,\text{d}\sigma_{\uu}(\phi)}. 
\end{split}
\end{align}
For experimental reasons, mainly to minimize 
effects of the \hermes spectrometer 
acceptance in $\phi$, the Fourier amplitudes 
associated with the asymmetry (\ref{eq1}) 
were extracted instead, e.g.,
\begin{align}\label{eq1d}
A_{\ut}^{\sin(\phi-\phi_S)}=
\frac{1}{4\pi^2}
\int\!\text{d}\phi \text{d}\phi_S \,\sin(\phi-\phi_S)\,
\frac{\text{d}\sigma_{\ut}(\phi,\phi_S)}{\text{d}\sigma_{\uu}(\phi)}. 
\end{align}
Similar equations hold for the other five amplitudes. 
These amplitudes embody all the essential information that could 
also be extracted from (\ref{eq1a}). For small (or zero) values of 
$\langle\cos\phi\rangle_{\uu}$ and $\langle\cos(2\phi)\rangle_{\uu}$, 
the amplitude in (\ref{eq1d}) corresponds to the one in (\ref{eq1b}).

The set of six Fourier amplitudes of the asymmetry
was obtained from the observed $\pi^+$ event sample 
using a maximum likelihood fit.  
The distribution of events was parameterized by the 
probability density function 
$\mathcal{N}_{\text{par}}$ defined as 
\begin{align}\label{eq3}
\mathcal{N}_{\text{par}}(P_\text{T},\phi,
\phi_S;\text{\boldmath$\eta$}_{\ut}) 
=1+P_\text{T}\,\mathcal{A}_{\ut}(\phi,\phi_S;
\text{\boldmath$\eta$}_{\ut}),
\end{align}
where
\begin{align}\label{eq2}
\begin{split}
&\mathcal{A}_{\ut}(\phi,\phi_S;\text{\boldmath$\eta$}_{\ut}) \\
&=A_{\ut}^{\sin(\phi-\phi_S)}\sin(\phi-\phi_S) 
+A_{\ut}^{\sin(\phi+\phi_S)}\sin(\phi+\phi_S) \\
&+A_{\ut}^{\sin\phi_S}\sin\phi_S 
+A_{\ut}^{\sin(2\phi-\phi_S)}\sin(2\phi-\phi_S) \\
&+A_{\ut}^{\sin(3\phi-\phi_S)}\sin(3\phi-\phi_S) 
+A_{\ut}^{\sin(2\phi+\phi_S)}\sin(2\phi+\phi_S). 
\end{split}
\end{align} 
Here, $\text{\boldmath$\eta$}_{\ut}$ represents the set of 
six Fourier amplitudes of the sine-modulation terms in (\ref{eq1d}).

Within the maximum likelihood scheme~\cite{Barlow},  
the logarithm of the 
likelihood function to be minimized is taken as 
\begin{align}\label{eq4}
\begin{split}
&\mathcal{L}(P_\text{T}^i,\phi^i,\phi_S^i;\text{\boldmath$\eta$}_{\ut})\\
&=-\sum_{i=1}^{N_{\pi^+}} \ln[1+P_\text{T}^i\,\mathcal{A}_{\ut}
(\phi^i,\phi_S^i;\text{\boldmath$\eta$}_{\ut})], 
\end{split}
\end{align}
where $N_{\pi^+}=N_{\pi^+}^\uparrow+N_{\pi^+}^\downarrow$ 
is the total number of events in the selected data sample, and

The raw results from the likelihood minimization of (\ref{eq4}) 
were corrected for background contributions in order to estimate the true 
results for exclusive $\pi^+$ production: 
\begin{align}\label{eq7}
A_\text{t}=\frac{A_\text{r}-b\,A_\text{b}}{1-b}. 
\end{align}
Here, $A_\text{r}$ stands for 
one of the six Fourier amplitudes in $\text{\boldmath$\eta$}_{\ut}$ 
(see (\ref{eq2}), (\ref{eq4})),  
$b$ and $A_\text{b}$ for the fractional contribution and corresponding 
Fourier amplitude of the background, 
and $A_\text{t}$ for the resulting true amplitude.
The background fraction is 
\begin{align}\label{eq6}
b=\frac{N_{\pi^+}-N_{\pi^+}^{\text{excl}}}{N_{\pi^+}},
\end{align}
where $N_{\pi^+}^{\text{excl}}$ is the number of exclusive events 
in the selected data sample. 

The following analysis was performed to estimate the quantities 
in (\ref{eq7}). 
As the recoiling neutron in the process $e p^{\uparrow}\to e n \pi^+$ 
was not detected, 
the sample of ``exclusive'' events was selected by requiring that 
the squared missing mass $M_X^2$ of the reaction 
$e p^{\uparrow}\to e \pi^+ X ~$ 
corresponds to the squared neutron mass $M_n^2$. 
The exclusive $\pi^+$ channel could not be completely 
separated from the channels with final states $\pi^+ + X$ 
(defined as background channels for $X\ne n$)
in which the $\pi^+$ originates, 
e.g., from neutral-meson (mainly $\rho^0$) decays, 
semi-inclusive processes, or  
nucleon resonance production, as their $M_X^2$ values were 
smeared into the region around $M_n^2$ due to the experimental 
resolution. 
These background events were subtracted from $N_{\pi^+}$ 
following the method briefly outlined below, and 
previously employed in the analysis of the exclusive $\pi^+$ 
cross section~\cite{hermes:xsection}.  
The exclusive $\pi^+$ yield was obtained by subtracting the 
yield difference  
$(N_{\pi^+} - N_{\pi^-})$ 
of the {\sc Pythia}~\cite{PYTHIA6} Monte Carlo simulation 
from that of the data, with both differences being independently absolutely 
normalized: 
\begin{align}\label{eq5}
N_{\pi^+}^{\text{excl}}=
(N_{\pi^+}-N_{\pi^-})^{\text{data}}
-(N_{\pi^+}-N_{\pi^-})^{\text{{\sc Pythia}}}.
\end{align} 
The {\sc Pythia} generator was used 
in conjunction with a set of 
{\sc Jetset}~\cite{Sjo94} fragmentation parameters 
that had previously been adjusted to reproduce 
exclusive vector meson production data~\cite{Patty} 
and multiplicity distributions~\cite{Hil05} 
observed by {\sc Hermes}. 
Exclusive production of single pions is absent in {\sc Pythia}. 
Note that exclusive $\pi^-$ mesons cannot be produced on protons. 
The constraint on the invariant mass of the initial 
photon-nucleon system $W^2>10\,\mathrm{GeV}^2$ was applied,  
and the pion 
momentum was required to satisfy $7\,\mathrm{GeV}<p<15\,\mathrm{GeV}$. 
Both conditions, applied to the data and the {\sc Pythia} yields,   
allowed for a better description of the data by the 
{\sc Pythia} Monte Carlo simulation for 
values of $M_X^2$ outside the region corresponding to exclusive 
$\pi^+$ production.
The resulting $M_X^2$ distribution  
of $N_{\pi^+}^{\text{excl}}$ 
and its resolution of $0.7\,\mathrm{GeV^2}$ were found to be 
consistent with that of a Monte Carlo sample of exclusive $\pi^+$ 
events normalized to the data 
(including radiative effects)~\cite{hermes:xsection}.

An ``exclusive region'' in $M_X^2$ was defined by requiring 
$-1.2\,\mathrm{GeV^2}<M_X^2<1.2\,\mathrm{GeV^2}$. 
The lower limit corresponds to three times the resolution of 
$M_X^2$, while the upper limit was set in order to minimize 
the (quadratically) combined statistical and systematic uncertainties 
of the extracted Fourier amplitudes. 
A relative systematic uncertainty of $20\%$ 
was assigned to $N_{\pi^+}^{\text{excl}}$, which corresponds to the largest 
data-to-\pythia discrepancy outside 
of the exclusive region~\cite{hermes:xsection}.  
As the $M_X^2$ spectrum of the positron-beam data is found to be 
shifted by approximately $0.16\,\mathrm{GeV^2}$ towards higher 
values relative to that of the electron-beam data, the exclusive region 
for the positron data is shifted accordingly. One quarter of the 
effect of this shift on the results presented below is assigned 
as a contribution to the systematic uncertainty.  

The values of $A_\text{r}$ and $b$ in (\ref{eq7}) are measured 
in the exclusive region. 
As the background originates from resolution smearing 
of events occurring at higher missing mass, $A_\text{b}$ in (\ref{eq7}) 
was assumed to be equal to 
the Fourier amplitude measured in
the $M_X^2$ region between $1.9\,\mathrm{GeV}^2$ and $3.3\,\mathrm{GeV}^2$ 
where the contribution of exclusive $\pi^+$ events is negligible. 
In that region $A_\text{b}$ was found to vary smoothly, with values   
smaller than $\pm0.1$, except for 
the $\sin\phi_S$ modulation for which 
it amounts on average to 
$(0.25\pm0.04)$. 
In order to account for a possible variation of $A_\text{b}$ 
with $M_X^2$ in the exclusive region, one half of the difference between
$A_\text{t}$ and $A_\text{r}$ is conservatively assigned as a contribution 
to the systematic uncertainty of $A_\text{t}$.

The values of $t^\prime$ were calculated from the measurement of
the four-momenta of the scattered lepton and produced pion by setting 
$M_X=M_n$, which improved the $t^\prime$-resolution by a factor of two. 
The kinematic range that contains the  
events used in the subsequent analysis is defined by the following 
requirements on the variables: 
$-t^\prime<0.7\,\mathrm{GeV}^2$, 
$0.03<x_\text{B}<0.35$, and $1\,\mathrm{GeV}^2<Q^2<10\,\mathrm{GeV}^2$. 
The mean $W^2$ value of the data is $16\,\mathrm{GeV}^2$. 

The dominant sources of systematic uncertainty  
are associated with the background subtraction and correction, 
and the observed relative shift of the $M_X^2$ distributions 
between positron and electron data. The contributions due to 
the residual beam polarization of $0.02\pm0.03$, 
the corresponding beam-spin asymmetry~\cite{Hermes2002}, 
and the charged-track curvature in the 
transverse field of the target magnet, 
are found to be negligible.
All these contributions, except for the target polarization 
scale uncertainty of $8.2\%$, 
are added in quadrature to yield the total systematic uncertainty.  
In addition, an estimate of the 
combined contribution to the experimental uncertainty
from resolution smearing, acceptance, kinematic bin width, and effects 
from the detector alignment with respect to the beam 
is determined using Monte Carlo 
simulation based on the GPD model~\cite{GK2009} 
for the $\sin(\phi-\phi_S)$ Fourier amplitude only.    
The difference between the amplitude extracted 
from the Monte Carlo sample and the corresponding 
model prediction calculated at the average kinematic values 
of the Monte Carlo sample 
is added in quadrature 
to the total systematic uncertainty of $A_{\ut}^{\sin(\phi-\phi_S)}$. 
The largest experimental uncertainties 
are those due to detector acceptance and kinematic bin width, and the 
determination of the target polarization. 

\begin{figure}[t]
\centering
\includegraphics[width=0.48\textwidth]{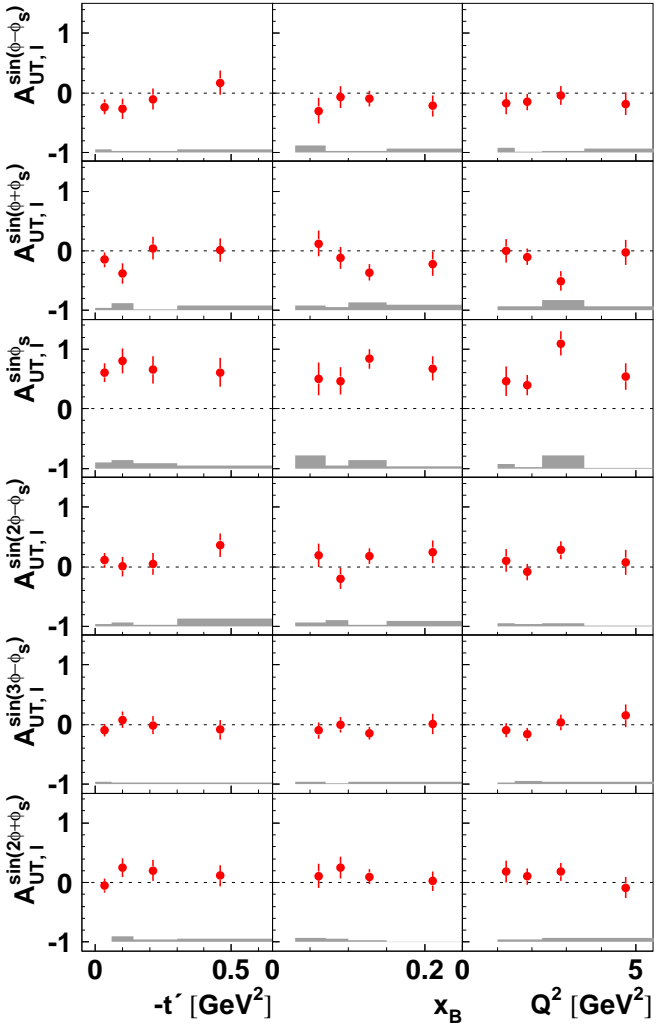}
\caption{
The set of six Fourier amplitudes 
($A_{\ut}$) describing the 
sine modulations of the single-spin   
azimuthal asymmetry for unpolarized (U) beam and 
transverse (T) target polarization, for the exclusive event sample. 
The error bars (bands) represent 
the statistical (systematic) uncertainties. The results receive 
an additional $8.2\%$ scale uncertainty 
corresponding to the target polarization uncertainty.
}
\label{ttsa}
\end{figure}

Figure~\ref{ttsa} shows the extracted Fourier amplitudes 
as a function of $-t^\prime$, $x_\text{B}$, and $Q^2$. 
For this measurement the average values of the kinematic variables are 
$\langle  -t^\prime\rangle=0.18\,\mathrm{GeV^2}$,
$\langle   x_\text{B}\rangle=0.13$, and
$\langle Q^2\rangle$=$2.38\,\mathrm{GeV^2}$. 
The background fraction $b$ 
varies between $(54\pm6)\%$ and $(62\pm5)\%$
in the various kinematic bins. 
As $x_\text{B}$ and $\langle Q^2\rangle$ are correlated 
the average values of $Q^2$ 
vary in the four $x_\text{B}$ bins, namely,  
$\langle Q^2\rangle=$ $1.24$, $1.57$, $2.24$, 
$3.91\,\mathrm{GeV}^2$. Analogously, 
the average values of $x_B$ vary in the four $Q^2$ bins, 
$\langle x_\text{B}\rangle=$ $0.07$, $0.11$, $0.15$, $0.23$. 
A separation of the contributions from longitudinal and transverse 
virtual photons to the Fourier amplitudes 
was not possible without measurements with different beam energies.

The six Fourier amplitudes shown in Fig.~\ref{ttsa} correspond 
to the following combinations of photoabsorption cross sections and 
interference terms $\sigma_{mn}^{ij}$ for photon helicities 
$m,n=0,\pm1$ and proton-spin projections 
$i,j=\pm\frac{1}{2}$~\cite{DiehlSapeta}:  
\begin{align}\label{eq2a}
\begin{split}
\langle\sin(\phi-\phi_S)\rangle_{\ut}
&\propto 
[\cos\theta\,\mathrm{Im}\,(\sigma_{++}^{+-}+\varepsilon\sigma_{00}^{+-}) \\
+ &\frac{1}{2}\sin\theta\sqrt{\varepsilon(1+\varepsilon)}
\mathrm{Im}\,(\sigma_{+0}^{++}-\sigma_{+0}^{--})],
\end{split}
\end{align}
\begin{align}
\begin{split}
\langle\sin(\phi+\phi_S)\rangle_{\ut} 
&\propto 
[\frac{1}{2}\cos\theta\,\varepsilon\,\mathrm{Im}\,\sigma_{+-}^{+-} \\
+ &\frac{1}{2}\sin\theta\sqrt{\varepsilon(1+\varepsilon)}
\mathrm{Im}\,(\sigma_{+0}^{++}-\sigma_{+0}^{--})],
\end{split}
\end{align}
\begin{align}
\begin{split}
\langle\sin\phi_S\rangle_{\ut} 
\propto 
[\cos\theta\sqrt{\varepsilon(1+\varepsilon)}
\mathrm{Im}\,\sigma_{+0}^{+-}],
\end{split}
\end{align}
\begin{align}
\begin{split}
\langle\sin(2\phi-\phi_S)\rangle_{\ut} 
&\propto 
[\cos{\theta}\sqrt{\varepsilon(1+\varepsilon)}\mathrm{Im}\,\sigma_{+0}^{-+}\\
+ &\frac{1}{2}\sin\theta\,\varepsilon
\mathrm{Im}\,\sigma_{+-}^{++}],
\end{split}
\end{align}
\begin{align}
\begin{split}
\langle\sin(3\phi-\phi_S)\rangle_{\ut} 
\propto 
[\frac{1}{2}\cos\theta\,\varepsilon\,
\mathrm{Im}\,\sigma_{+-}^{-+}],
\end{split}
\end{align}
\begin{align}
\begin{split}
\langle\sin(2\phi+\phi_S)\rangle_{\ut}  
\propto 
[\frac{1}{2}\sin\theta\,\varepsilon\,
\mathrm{Im}\,\sigma_{+-}^{++}],
\end{split}
\end{align}
where $\varepsilon$ is the virtual-photon polarization parameter, 
and $\theta$ is the angle between the 
beam and the virtual-photon direction. 
Note that in the analysis presented here there is an integration over a 
range in \(\theta\), with \(\cos\theta \approx 1\) and \(0.04\le\sin\theta\le0.15\).
 At leading twist, only $\langle\sin(\phi-\phi_S)\rangle_{\ut}$ 
receives a contribution from 
only longitudinal virtual photons via $\sigma_{00}^{+-}$,  
while the other Fourier amplitudes are expected to be 
suppressed~\cite{Collins} 
by at least one power of $1/Q$ due to interference between 
contributions from longitudinal and transverse virtual photons, 
and by $1/Q^2$ due to terms involving only transverse virtual photons. 

Most of the Fourier amplitudes 
shown in Fig.~\ref{ttsa} are small or consistent with zero, except 
$A_{\ut}^{\sin\phi_S}$. This amplitude 
is found to be large and positive indicating a significant contribution 
from the transverse-to-longitudinal helicity transition of 
the virtual photon, i.e., 
\begin{align}
\begin{split}
A_{\ut}^{\sin\phi_S} 
\propto \sigma_{+0}^{+-}=
\sum_{\nu^\prime}\mathcal{M}_{0\nu^\prime ++}^*\,\mathcal{M}_{0\nu^\prime 0-}\\
=\mathcal{M}_{0+++}^*\,\mathcal{M}_{0+0-}
+\mathcal{M}_{0-++}^*\,\mathcal{M}_{0-0-},
\end{split}
\end{align}
where $\mathcal{M}_{\mu^\prime\nu^\prime\mu\nu}$ are helicity amplitudes 
with $\mu^\prime$ ($\mu$) and $\nu^\prime$ ($\nu$) denoting the helicities 
of the pion (virtual photon) and the neutron (proton), respectively.  
These amplitudes are proportional to 
$\sqrt{-t^\prime}^{|\mu-\nu-\mu^\prime+\nu^\prime|}$. 
In the framework of GPDs, the
amplitude $\mathcal{M}_{0-++}$
is associated at leading twist with virtual-photon helicity 
flip in the $t$-channel~\cite{DiehlSapeta}, which is 
proportional to 
$\sqrt{-t^\prime}$ and hence is expected to vanish 
for $-t^\prime\rightarrow 0$. 
However, among higher-twist contributions 
the one that involves the 
parton-helicity-flip GPDs $H_T$ and $\widetilde{H}_T$  
need not vanish at small values of $|t^\prime|$. 
Moreover, in the more general framework of helicity amplitudes and 
the Regge model, 
$A_{\ut}^{\sin\phi_S}$ 
receives contributions from natural 
and unnatural-parity exchange~\cite{Diehl2008,GK2009}, 
which allow it 
to remain constant as a function of $-t^\prime$, 
as the data in Fig.~\ref{ttsa} suggest. 
Lack of parameterizations of $\sigma_{mn}^{ij}$ involving 
transverse virtual photons does not allow further interpretation 
of the corresponding Fourier amplitudes. Any model that describes 
exclusive pion 
production will need to describe not only the leading-twist Fourier amplitude, 
but also the other
contributions to the target-spin azimuthal asymmetry.

Of special interest in the present measurement 
is the Fourier amplitude 
$A_{\ut}^{\sin(\phi-\phi_S)}$
in case of production by longitudinal photons, 
which can be compared with GPD models. 
It is related to the 
parton-helicity-conserving part of the scattering process and
is sensitive to the interference between 
$\widetilde{\mathcal{H}}$ and 
$\widetilde{\mathcal{E}}$~\cite{Pobylitsa,DiehlKugler}:
\begin{align}
\begin{split}
A_{\ut}^{\sin(\phi-\phi_S)} = &
-\frac{\sqrt{-t^\prime}}{M_p} \\
\times & \frac{\xi\sqrt{1-\xi^2}
\,\mathrm{Im}(\widetilde{\mathcal{E}}^*\widetilde{\mathcal{H}})}
{(1-\xi^2)\widetilde{\mathcal{H}}^2
-\frac{t\xi^2}{4M_p^2}\,\widetilde{\mathcal{E}}^2
-2\xi^2\,\mathrm{Re}(\widetilde{\mathcal{E}}^*\widetilde{\mathcal{H}})},
\end{split}
\end{align}
where the 
transition form factors  
$\widetilde{\mathcal{H}}$ and $\widetilde{\mathcal{E}}$  
denote convolutions of hard scattering kernels and the pion 
distribution amplitude with the GPDs  
$\widetilde H$ and $\widetilde E$, respectively. 
Note that in the models described below terms 
proportional to the $\cos\phi$ and 
$\cos(2\phi)$ modulation 
of the spin-averaged cross section are not included.   
In the measurement presented here these terms 
are not known, although they nonetheless  
contribute to the values of the extracted Fourier amplitudes.

\begin{figure}[t]
\centering
\includegraphics[width=0.48\textwidth]{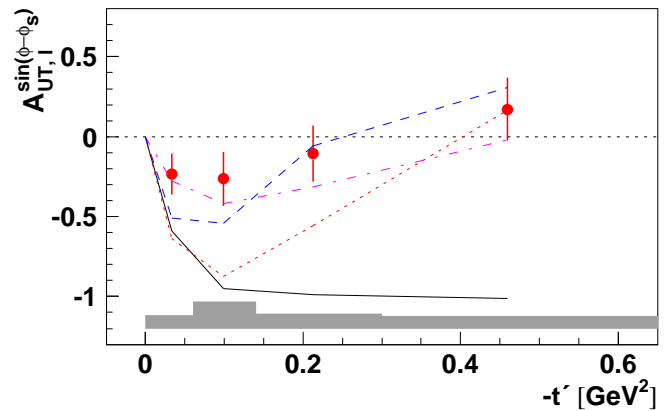}
\caption{
Model predictions for the $\sin(\phi-\phi_S)$ Fourier amplitude
as a function of $-t^\prime$. 
The curves represents predictions of GPD-model calculations. 
The full circles show the values of $A_{\ut}^{\sin(\phi-\phi_S)}$ taken 
from Fig.~\ref{ttsa}.
The error bars (bands) represent the statistical (systematic) uncertainties. 
See text for details.
}
\label{tdep}
\end{figure}

Figure~\ref{tdep} shows the extracted Fourier amplitude 
$A_{\ut}^{\sin(\phi-\phi_S)}$ as a function of $-t^\prime$ 
in more detail. The solid and dotted curves represent the leading-twist, 
leading-order in $\alpha_s$ calculations of 
this amplitude for longitudinal virtual photons  
using two variants of the GPD model of~\cite{BM2009}.
The modelling of the GPD $\widetilde{E}$ relies here, 
even at larger values of $-t$, 
on the dominance of the pion pole $1/(m_\pi^2-t)$ 
in the pion exchange amplitude, 
with $m_\pi$ the pion mass. Then   
$\widetilde{\mathcal{E}}$ is real and positive, and the value of 
$A_{\ut}^{\sin(\phi-\phi_S)}$ is typically predicted to be large and 
negative, while it must sharply vanish at the kinematic boundary $-t^\prime=0$ 
(see solid curve). The data qualitatively disagree with such a 
simplified GPD model.
The ``Regge-ized'' variant of the GPD-$\widetilde{E}$ model~\cite{BM2009}, 
containing more than only a pion $t$-channel exchange, 
results in the dash-dotted curve. In such a model the asymmetry 
can become positive at larger values of $-t^\prime$, caused by a 
negative real part in $\widetilde{\mathcal{E}}$.  The dash-dotted curve 
arises from an alternative GPD approach~\cite{KMPK08}, 
in which the imaginary part of 
$\widetilde{\mathcal{H}}$ becomes negative while the real part 
of $\widetilde{\mathcal{E}}$ remains positive 
at larger values of $-t^\prime$.  

An attempt to evaluate the complete set of Fourier amplitudes~(\ref{eq2}),  
and in particular the value of $A_{\ut}^{\sin(\phi-\phi_S)}$,  
is presented in~\cite{GK2009}. 
In this model, the GPDs are calculated in a similar way as 
in the models~\cite{Belitsky2001,VGG}, except that 
the experimental value of 
the pion form factor $F_\pi$ is used. 
Here a large non-pole contribution from $\widetilde{E}$ 
over-compensates the pion-pole contribution leading to the 
zero-crossing  
behavior of the amplitude as a function of $-t^\prime$    
(see dashed curve in Fig.~\ref{tdep}).  
This model appears to be qualitatively in agreement with the data. 
However, within the large experimental uncertainty 
$A_{\ut}^{\sin(\phi-\phi_S)}$
is also consistent with zero. A vanishing Fourier amplitude 
in this model 
implies the dominance (due to the pion pole) of $\widetilde E$ 
over $\widetilde H$ at low $-t^\prime$. 
This is in agreement with the recent \hermes measurement of the 
exclusive $\pi^+$ cross section~\cite{hermes:xsection}, which 
is well described at $-t^\prime=0.1\,\mathrm{GeV^2}$ by a 
GPD model~\cite{VGG} based only on $\widetilde E$ while neglecting 
the contribution of $\widetilde H$. 

In summary, the Fourier amplitudes of the 
single-spin azimuthal asymmetry are 
measured in exclusive electroproduction 
of $\pi^+$ mesons on transversely polarized protons, 
for the first time. Within the experimental uncertainties 
the amplitude of the $\sin(\phi-\phi_S)$ 
modulation is found to be consistent with zero, thus  
excluding a pure pion-pole contribution to the GPD $\widetilde E$ 
in leading-twist calculations.  
This could also be an indication for the dominance  
of $\widetilde E$ over the GPD $\widetilde H$ at low $-t^\prime$.   
The observed amplitude of the $\sin\phi_S$ modulation is large and 
positive which implies the presence of a sizeable 
interference between contributions from longitudinal and transverse 
virtual photons. 
A next-to-leading twist calculation as well as 
knowledge of the contributions from transverse photons 
and their interference with longitudinal photons 
are required for a description of the measurements.

We gratefully acknowledge the \desy management for its support and the staff
at \desy and the collaborating institutions for their significant effort.
This work was supported by the IWT and FWO-Flanders, Belgium;
the Natural Sciences and Engineering Research Council of Canada;
the National Natural Science Foundation of China;
the Alexander von Humboldt Stiftung;
the German Bundesministerium f\"ur Bildung und Forschung (BMBF);
the Deutsche Forschungsgemeinschaft (DFG);
the Italian Istituto Nazionale di Fisica Nucleare (INFN);
the MEXT, JSPS, and COE21 of Japan;
the Dutch Foundation for Fundamenteel Onderzoek der Materie (FOM);
the U. K. Engineering and Physical Sciences Research Council, the
Particle Physics and Astronomy Research Council and the
Scottish Universities Physics Alliance;
the U. S. Department of Energy (DOE) and the National Science Foundation (NSF);
the Russian Academy of Science and the Russian Federal Agency for 
Science and Innovations;
the Ministry of Trade and Economical Development and the Ministry
of Education and Science of Armenia;
and the European Community-Research Infrastructure Activity under the
FP6 ''Structuring the European Research Area'' program
(HadronPhysics, contract number RII3-CT-2004-506078).

\end{document}

%% file: authors.tex










\def\groupargonne{\affiliation{Physics Division, Argonne National Laboratory, Argonne, Illinois 60439-4843, USA}}
\def\groupbari{\affiliation{Istituto Nazionale di Fisica Nucleare, Sezione di Bari, 70124 Bari, Italy}}
\def\groupbeijing{\affiliation{School of Physics, Peking University, Beijing 100871, China}}
\def\groupcolorado{\affiliation{Nuclear Physics Laboratory, University of Colorado, Boulder, Colorado 80309-0390, USA}}
\def\groupdesy{\affiliation{DESY, 22603 Hamburg, Germany}}
\def\groupzeuthen{\affiliation{DESY, 15738 Zeuthen, Germany}}
\def\groupdubna{\affiliation{Joint Institute for Nuclear Research, 141980 Dubna, Russia}}
\def\grouperlangen{\affiliation{Physikalisches Institut, Universit\"at Erlangen-N\"urnberg, 91058 Erlangen, Germany}}
\def\groupferrara{\affiliation{Istituto Nazionale di Fisica Nucleare, Sezione di Ferrara and Dipartimento di Fisica, Universit\`a di Ferrara, 44100 Ferrara, Italy}}
\def\groupfrascati{\affiliation{Istituto Nazionale di Fisica Nucleare, Laboratori Nazionali di Frascati, 00044 Frascati, Italy}}
\def\groupgent{\affiliation{Department of Subatomic and Radiation Physics, University of Gent, 9000 Gent, Belgium}}
\def\groupgiessen{\affiliation{Physikalisches Institut, Universit\"at Gie{\ss}en, 35392 Gie{\ss}en, Germany}}
\def\groupglasgow{\affiliation{Department of Physics and Astronomy, University of Glasgow, Glasgow G12 8QQ, United Kingdom}}
\def\groupillinois{\affiliation{Department of Physics, University of Illinois, Urbana, Illinois 61801-3080, USA}}
\def\groupmichigan{\affiliation{Randall Laboratory of Physics, University of Michigan, Ann Arbor, Michigan 48109-1040, USA }}
\def\groupmoscow{\affiliation{Lebedev Physical Institute, 117924 Moscow, Russia}}
\def\groupnikhef{\affiliation{National Institute for Subatomic Physics (Nikhef), 1009 DB Amsterdam, The Netherlands}}
\def\groupstpetersburg{\affiliation{Petersburg Nuclear Physics Institute, Gatchina, Leningrad region, 188300 Russia}}
\def\groupprotvino{\affiliation{Institute for High Energy Physics, Protvino, Moscow region, 142281 Russia}}
\def\groupregensburg{\affiliation{Institut f\"ur Theoretische Physik, Universit\"at Regensburg, 93040 Regensburg, Germany}}
\def\grouprome{\affiliation{Istituto Nazionale di Fisica Nucleare, Sezione Roma 1, Gruppo Sanit\`a and Physics Laboratory, Istituto Superiore di Sanit\`a, 00161 Roma, Italy}}
\def\grouptriumf{\affiliation{TRIUMF, Vancouver, British Columbia V6T 2A3, Canada}}
\def\grouptokyo{\affiliation{Department of Physics, Tokyo Institute of Technology, Tokyo 152, Japan}}
\def\groupamsterdam{\affiliation{Department of Physics and Astronomy, Vrije Universiteit, 1081 HV Amsterdam, The Netherlands}}
\def\groupwarsaw{\affiliation{Andrzej Soltan Institute for Nuclear Studies, 00-689 Warsaw, Poland}}
\def\groupyerevan{\affiliation{Yerevan Physics Institute, 375036 Yerevan, Armenia}}
\def\groupnone{\noaffiliation}


\groupargonne
\groupbari
\groupbeijing
\groupcolorado
\groupdesy
\groupzeuthen
\groupdubna
\grouperlangen
\groupferrara
\groupfrascati
\groupgent
\groupgiessen
\groupglasgow
\groupillinois
\groupmichigan
\groupmoscow
\groupnikhef
\groupstpetersburg
\groupprotvino
\groupregensburg
\grouprome
\grouptriumf
\grouptokyo
\groupamsterdam
\groupwarsaw
\groupyerevan


\author{A.~Airapetian}  \groupgiessen \groupmichigan
\author{N.~Akopov}  \groupyerevan
\author{Z.~Akopov}  \groupdesy
\author{E.C.~Aschenauer}  \groupzeuthen
\author{W.~Augustyniak}  \groupwarsaw
\author{A.~Avetissian}  \groupyerevan
\author{E.~Avetisyan}  \groupdesy
\author{B.~Ball}  \groupmichigan
\author{S.~Belostotski}  \groupstpetersburg
\author{N.~Bianchi}  \groupfrascati
% ...

\collaboration{The HERMES Collaboration} \noaffiliation




